\documentclass{article}
\usepackage{smc2021}
\usepackage[caption=false, font=footnotesize]{subfig}
\usepackage{paralist}
\usepackage[figure,table]{hypcap}

\usepackage{amsmath}

\usepackage{arabtex}
\usepackage[LFE,LAE,LGR,T2A,T1]{fontenc}
\usepackage[main=english]{babel}
\usepackage{xcolor}
\usepackage[colorinlistoftodos]{todonotes}

\def\papertitle{Signal Representations for Synthesizing Audio Textures with Generative Adversarial Networks}
\author[1]{\mbox{\firstname{Chitralekha}\lastname{Gupta}\email{chitralekha@nus.edu.sg}\orcid{0000-0003-1350-9095}}}
\author[1]{\mbox{\firstname{Purnima}\lastname{Kamath}\email{purnima.kamath@u.nus.edu}\orcid{0000-0003-0351-6574}}}
\author[1]{\mbox{\firstname{Lonce}\lastname{Wyse}\email{lonce.wyse@nus.edu.sg}\orcid{ 0000-0002-9200-1048}}}

\affil[1]{\institution{National University of Singapore}\country{Singapore}\affiliationtype{University}}

\completesetup

\title{\papertitle}
\begin{document}
	\capstartfalse
	\maketitle
	\capstarttrue

	\begin{abstract}
		 Generative Adversarial Networks (GANs) currently achieve the state-of-the-art sound synthesis quality for pitched musical instruments using a 2-channel spectrogram representation consisting of log magnitude and instantaneous frequency (the "IFSpectrogram"). Many other synthesis systems use representations derived from the magnitude spectra, and then depend on a backend component to invert the output magnitude spectrograms that generally result in audible artefacts associated with the inversion process. However, for signals that have closely-spaced frequency components such as non-pitched and other noisy sounds, training the GAN on the 2-channel IFSpectrogram representation offers no advantage over the magnitude spectra based representations. In this paper, we propose that training GANs on single-channel magnitude spectra, and using the Phase Gradient Heap Integration (PGHI) inversion algorithm is a better comprehensive approach for audio synthesis modeling of diverse signals that include  pitched, non-pitched, and dynamically complex sounds. We show that this method produces higher-quality output for wideband and noisy sounds, such as pops and chirps, compared to using the IFSpectrogram. Furthermore, the sound quality for pitched sounds is comparable to using the IFSpectrogram, even while using a simpler representation with half the memory requirements.
	\end{abstract}

	\section{Introduction}\label{sec:introduction}
	In recent years, GANs have achieved the state-of-the-art performance in neural audio synthesis, specifically for pitched musical instrument sounds \cite{engel2019gansynth,nistal2021comparing}. Engel et al.\cite{engel2019gansynth} showed that a progressively growing GAN \cite{karras2017progressive} can outperform strong WaveNet \cite{van2016wavenet} and WaveGAN \cite{donahue2018adversarial} baselines in the task of conditional musical instrument audio generation achieving comparable audio synthesis quality and faster generation time. Nistal et al.\cite{nistal2021comparing} further showed that a 2-channel input representation consisting of the magnitude and the instantaneous frequency (IF) of the Short-Time Fourier Transform (STFT) achieves the best synthesis results in this framework compared to other kinds of representations, such as Mel spectrogram, MFCC, and Constant-Q Transform. Estimation of IF, which is the derivative of the unwrapped phase with respect to time, provides comprehensive information about the phase of the signal when the audio is pitched, i.e.~has components that are clearly separated in frequency.
	Thus, a magnitude spectrogram combined with the estimated IF results in high-quality 
	reconstruction of the signal for pitched signals such as musical instruments. In broadband and noisy short duration signals, components are not separated in frequency, and neighboring frequency bins have complex and highly interdependent amplitude and phase relationships that are necessary for reconstruction and the representation is very sensitive to IF estimation errors.
	
	DrumGAN \cite{nistal2020drumgan} extended the work in \cite{nistal2021comparing} to various drum sounds, however the authors have notably not used the IF spectrogram that produce state-of-the-art quality for pitched sounds, but instead, use spectrograms of the real and imaginary parts from the STFT directly. They also use a set of perceptually correlated features more appropriate than pitch for conditioning the percussion sounds in the target data set.
	
	Pr\r{u}{\v{s}}a et al.~\cite{pruuvsa2017noniterative} proposed a non-iterative phase reconstruction algorithm called Phase Gradient Heap Integration (PGHI) that uses the mathematical relationship between the magnitude of Gaussian windowed STFT and the phase derivatives in time and frequency of the Fourier transform to reconstruct the phase using only the magnitude spectrogram. Marafioti et al.~\cite{marafioti2019adversarial} compared three different GAN  architectures, and showed that for a dataset consisting of spoken digits and piano music, the architecture using PGHI produced audio of objectively and perceptually higher quality than the other representations they compared based on  an aggregate set of different signal types. A direct comparison with GanSynth\cite{engel2019gansynth} which was being published at about the same time was also not included in their study.
	
	In this paper, we study and compare the state-of-the-art GanSynth with magnitude spectrogram+IF audio representation and reconstruction method and the PGHI method of representation and reconstruction for a systematically organized collection of audio textures such as pitched musical instruments, noisy pops, and chirps, spanning a range from pitched steady-state to broadband signals. We show that the PGHI method of reconstruction from GAN estimates is more robust for synthetic spectrograms and estimation errors for different kinds of input signals than the state-of-the-art magnitude+IF representation.
	This study contributes to the development of general and efficient representations for training GANs for complex audio texture synthesis.  
	\section{Audio Textures and Representations}
		\subsection{Audio Representations and Inversion Techniques}
	\label{sec:audiorep}
	Many algorithms learn to estimate the magnitude spectrogram and then use iterative methods such as Griffin-Lim to estimate the phase and reconstruct the time domain signal. However, these traditional methods of phase estimation and reconstruction are known to have perceptible artifacts in the reconstructed signal. Estimation of phase is difficult and prone to errors in part because artificial or manipulated images may not produce a real-valued time domain signal when inverted.
	
	Another way of representing phase is with instantaneous frequency. A sinusoidal wave with a constant frequency produces a phase, which when unwrapped grows linearly. The derivative of this unwrapped phase with respect to time remains constant and is equal to the angular difference between the frame stride and signal periodicity, and is commonly referred to as the instantaneous frequency (IF). The estimate of magnitude spectrogram and IF in frequency domain can be used to reconstruct a time domain signal by computing the unwrapped phase from the cumulative sum of IF across time axis, and computing an inverse Fourier transform. The state-of-the-art GANSynth framework \cite{engel2019gansynth,nistal2021comparing} estimates this 2-channel audio representation, i.e. log magnitude and IF, or IFSpectrogram. 
	Engel et al. hypothesized and showed that synthesized audio quality from the IFSpectrogram is robust to estimation errors for the NSynth dataset of pitched musical instrument audio while noting the importance of choosing analysis window sizes large enough to be primarily sensitive to a single frequency component.
Nistal et al.~\cite{nistal2021comparing} compared different audio representations such as waveform, complex spectrogram, melspectrogram, cqt spectrogram, and IFSpectrogram, and found that synthesis of the pitched musical instruments from the estimates of IFSpectrogram provides the best audio quality. However, to the best of our knowledge, IFSpectrogram method has not been tested and compared to other representations 
	for non-pitched and noisy sounds. 
	
We observe that whether converting  pitched instrument or noisy transient audio into IFSpectrogram representation, that resynthesizing produces a high quality audio output for both the kinds of sounds. However, if we add a small Gaussian noise to the IF channel (to simulate estimation error in IF) and then resynthesize, the perceptual quality of the pitched sounds is not affected as much as the quality of the noisy pop sounds. Audio examples of this simulation are presented here: \url{https://animatedsound.com/amt/listening_test_samples/#simulation}. This indicates that IFSpectrogram method may not be robust to manipulated and synthetic spectrograms or estimation errors for non-pitched and noisy sounds.
	
For a signal composed of sinusoidal components with constant frequencies, the phase grows linearly in time for all the frequency channels that have energy in the spectrogram. For these coefficients, the IF is constant and the local group delay (STFT phase derivative with respect to frequency) is zero. However, in case of an impulse train, the situation is reverse to that of sinusoidal components, wherein the phase derivative with respect to frequency axis will have more information than the IF as there is energy across almost all the frequency channels in the spectrogram, but the change of phase with respect to time exists only around the impulse events, and otherwise it is zero. Furthermore, for signals that have fast moving or closely spaced frequency components, IF does not capture the variability in the frequency direction. 

The Phase Gradient Heap Integration (PGHI) method \cite{pruuvsa2017noniterative} is a non-iterative phase estimation method that exploits the mathematical relationship between the time and frequency derivatives of log magnitude spectrogram with the phase gradients in frequency and time axes respectively. To provide a brief summary here, Pr\r{u}{\v{s}}a et al.\cite{pruuvsa2017noniterative} proved mathematically and experimentally that the derivative of phase along frequency axis $\phi_\omega(m,n)$ and, the derivative of phase along time axis $\phi_t(m,n)$ can be estimated solely from the time and frequency derivatives of log-magnitude of STFT ($\mathrm{s_{log}}_t$, $\mathrm{s_{log}}_\omega$) respectively computed with a Gaussian window, as \cite{prusa2017phase,Prusa2017},
\begin{equation}
	\begin{aligned}
	    \phi_\omega(m,n)=\frac{-\gamma}{2aM}(\mathrm{s_{log}}_t(m,n)) \\
	    \phi_t(m,n)=\frac{aM}{2\gamma}(\mathrm{s_{log}}_\omega(m,n))+2\pi am/M
	\end{aligned}
\end{equation}
where, $M$ is the number of frequency channels, $a$ is the hop size, and $\gamma$ is the time-frequency ratio of Gaussian window, which is recommended to be $aM/L$, $L$ being the length of the input signal in samples.
Redundancy between frames should be such that there is sufficient dependency between the values of the STFT to facilitate magnitude-only reconstruction. The recommended redundancy is $M/a\geq4$ \cite{marafioti2019adversarial}.
	
	This method also implements a numerical integration of these phase gradients such that integration is first performed along the prominent contours of the spectrogram in order to reduce accumulation of the error, and so on. This heap integration method to estimate phase from the phase gradients helped to make the synthesis robust to estimation errors and noise \cite{prusa2017phase,pruuvsa2017noniterative}.

Here we show that training a GAN on a single channel log magnitude spectrogram and using the PGHI algorithm for inversion of the estimated spectrogram to time-domain signal produces better audio quality for wideband, noisy, non-pitched or fast changing signals than when using the IFSepctrogram representation to train the state-of-the-art GAN for pitched data. Moreover, although the single-channel representation requires half the memory, the audio quality of the pitched sounds produced by PGHI is also comparable to that of IFSpectrogram. This is thus a general approach for audio synthesis using the state-of-the-art GAN that works for a variety of different sounds.
	\subsection{Audio Textures}
	\label{sec:audiotextures}
	Audio synthesis finds practical applications in creative sound design for music, film, and gaming, where creators are looking for sound effects suited to specific scenarios. Research in this field aims to learn a compact latent space of audio such that adjustments to these latent variables would help the creator search through a known space of sounds (eg. water drops and footsteps), parametrically control (eg. rate of water dripping) as well as explore new sounds in the spaces in between the known sounds\cite{donahue2018adversarial}. 

    Building upon generative adversarial image synthesis techniques, researchers exploring GAN techniques for neural audio synthesis have made significant progress in building frameworks for conditional as well as unconditional synthesis of a wide range of musical instrument timbres \cite{engel2019gansynth,nistal2021comparing}. These models are trained on NSynth dataset \cite{engel2017neural} that consists of notes from musical instruments across a range pitches, timbres, and volumes. Conditioning on pitch allows the network to learn natural timbre variation while providing musical control of notes for synthesis.
    The NSynth dataset provides a comprehensive representation of pitched sounds comprised primarily of well-separated harmonics. There has been some work on audio texture modeling for synthesis \cite{saint1995analysis, schwarz2011state, mcdermott2009sound} including deep learning approaches \cite{antognini2019audio}, but audio textures have received considerably less attention than  traditional musical sounds and speech.
    
\begin{figure*}
  \centering
  \mbox{
    \subfloat[Pitched Instrument (Piano)]{\includegraphics[width=0.27\linewidth]{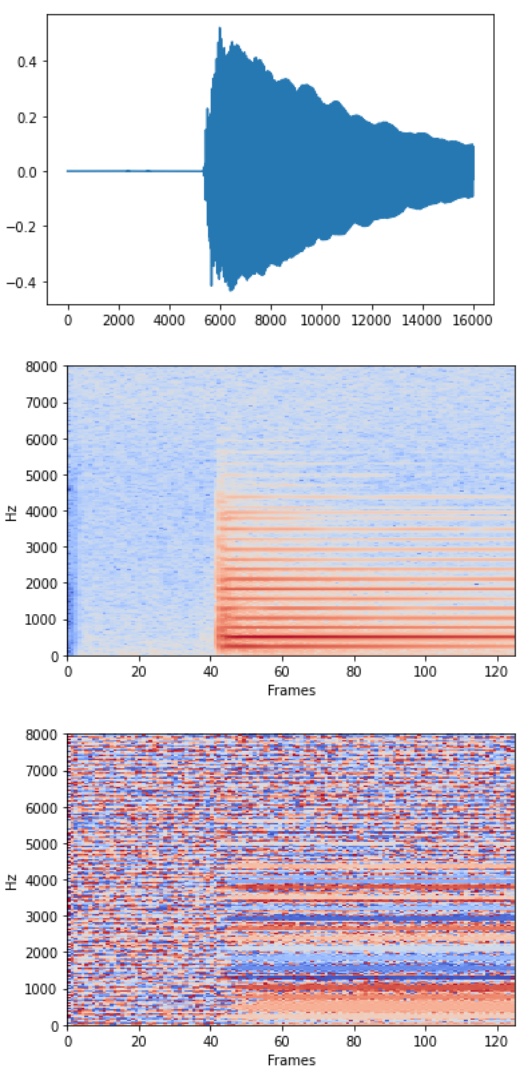}}\quad
    \subfloat[Pops]{\includegraphics[width=0.27\linewidth]{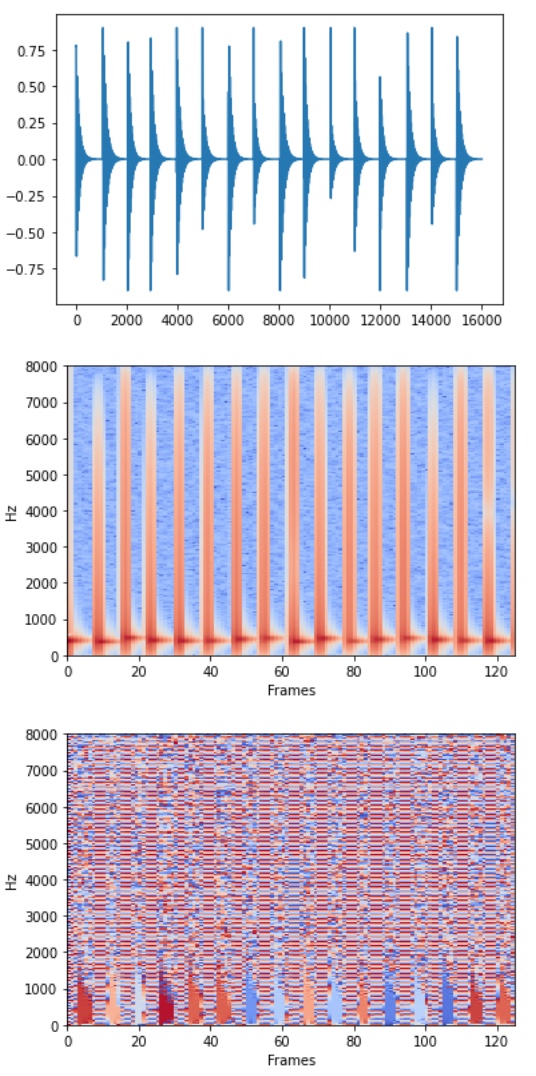}}\quad
    \subfloat[Chirps]{\includegraphics[width=0.27\linewidth]{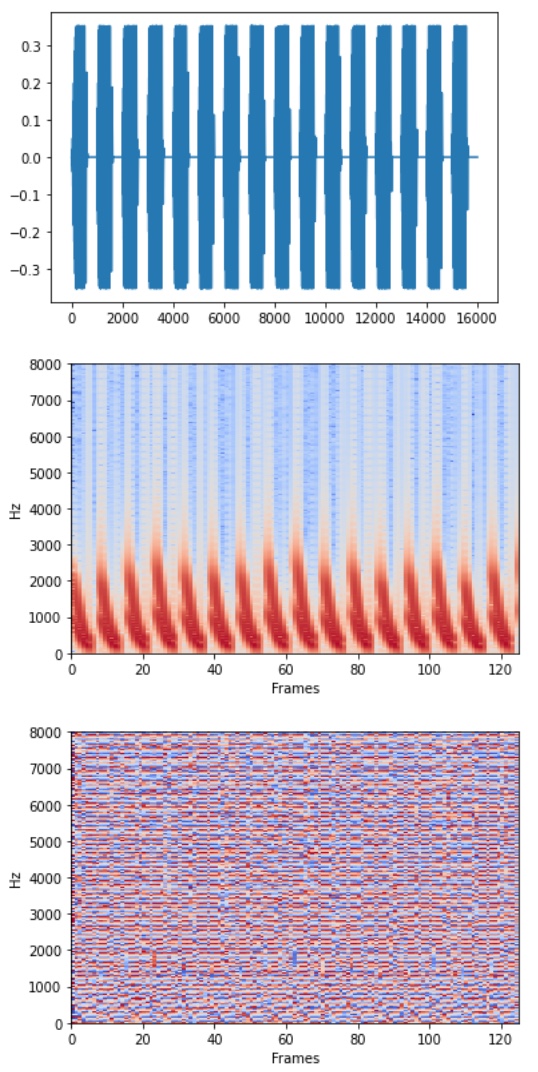}}
  }
  \caption{Examples of (a) a pitch instrument (piano), (b) Noise burst or pops, and (c) Frequency sweeps or chirps, with their respective audio waveform (top row), log magnitude spectrogram (middle row), and instantaneous frequency of unwrapped phase (bottom row) plots.}
  \label{fig:audio_textures}
\end{figure*}

    Sound textures \cite{saint1995analysis, wyse2020AIMusicTextures} have more timbral variation including wideband or noisy components, such as footsteps or motors, and a wide range of temporal structure not found in pitched instruments. Furthermore, there can be very fast-varying frequency components and pitches in sounds such as water dripping, and chirps. Thus we examine the performance of controlled audio synthesis techniques on trained networks using three types of sounds - pitched instruments, noise burst pops, and frequency sweep chirps, as shown in Figure \ref{fig:audio_textures}.
	
	\subsection{Conditional GAN architecture for audio synthesis}\label{sec:typeset_text}
	Parametrically controllable audio synthesis has also been an active field of research in recent years. Hsu et al.~\cite{hsu2018hierarchical} used hierarchical variational autoencoders (VAEs) for conditional or controlled speech generation. Similarly, Luo et al.~\cite{luo2019learning} learn separate latent distributions using VAEs to control the pitch and timbre of musical instrument sounds. Engel et al.~\cite{engel2017neural} conditioned a WaveNet-style autoregressive model to generate musical sounds, as well as interpolate between sounds to generate new sounds. The current state-of-the-art performance in conditional synthesis of audio is the GANSynth architecture \cite{engel2019gansynth} which introduces a progressively growing Wasserstein GAN for controlled music synthesis and is based on the IFSpectrogram representation \cite{nistal2021comparing}. Thus, we adopt this architecture with IFSpectrogram representation as our baseline. 
	
	\section{Experimental Details}
	\subsection{Audio Datasets}
	
	\subsubsection{Pitched Musical Instruments}
	We make use of the NSynth dataset \cite{engel2017neural}, that consists of approximately 300,000 single-note audios played by more than 1,000 different instruments. It contains labels for pitch, velocity, instrument type, acoustic qualities (acoustic or electronic), and more, although, for this particular work, we only make use of the pitch information as the conditional parameter. We use the same subset of this dataset as was used by Nistal et al.~\cite{nistal2021comparing}. It contains acoustic instruments from the brass, flutes, guitars, keyboards, and mallets families, and the audio samples are trimmed from 4 to 1 seconds and only consider samples with a MIDI pitch range from 44 to 70 (103.83 - 466.16 Hz). This yields a subset of approximately 22k audio files with balanced instrument class distribution.
	\subsubsection{Noisy Pops}
	On the other end of the spectrum of sounds we tested are \textit{pops.} A pop is a burst of noise filtered by a bandpass filter. We generated the pop textures with three parameters - rate (number of events per seconds), irregularity in the temporal distribution (using a Gaussian distribution around each evenly-spaced time value), and the center frequency of the bandpass filter. Rate ranges from 2 to 16 pops per second, center frequency ranges from 440 to 880 Hz (corresponding to midi pitch values 69 to 81), and irregularity described by a Gaussian distribution with a standard deviation ranging from 0.04 to 0.4. We generate 21 values for each of these 3 parameters, and five 1 second long audio clips of each combination, resulting in a total of 46,305 (21*21*21*5) audio files.
	\subsubsection{Chirps}
	In between the quality of the pitched sounds with relatively steady frequency components and the noisy pop sounds with sharp broadband transients are \textit{chirps}. A chirp is a signal in which the frequency increases or decreases quickly with time. The chirps were generated with two frequency components space by an octave, and were controlled with 5 parameters - irregularity in time (like the pops), chirp rate (2 to 16 chirps per second, 9 samples), frequency sweep range in octaves ([-3. -1, 1, 3]), event duration (5 linearly spaced samples in [.02, .2]), and center frequency (9 linearly space samples in musical pitch space between 440 and 880 Hz). We generate 5 variations of each parameter (different due to the statistical distribution of events in time) resulting in a total of 40,500 (5*9*4*5*9*5) audio files of 1 second each.

	\subsection{GAN architecture}
\begin{figure}
	    \centering
  \subfloat[]{\includegraphics[width=\linewidth]{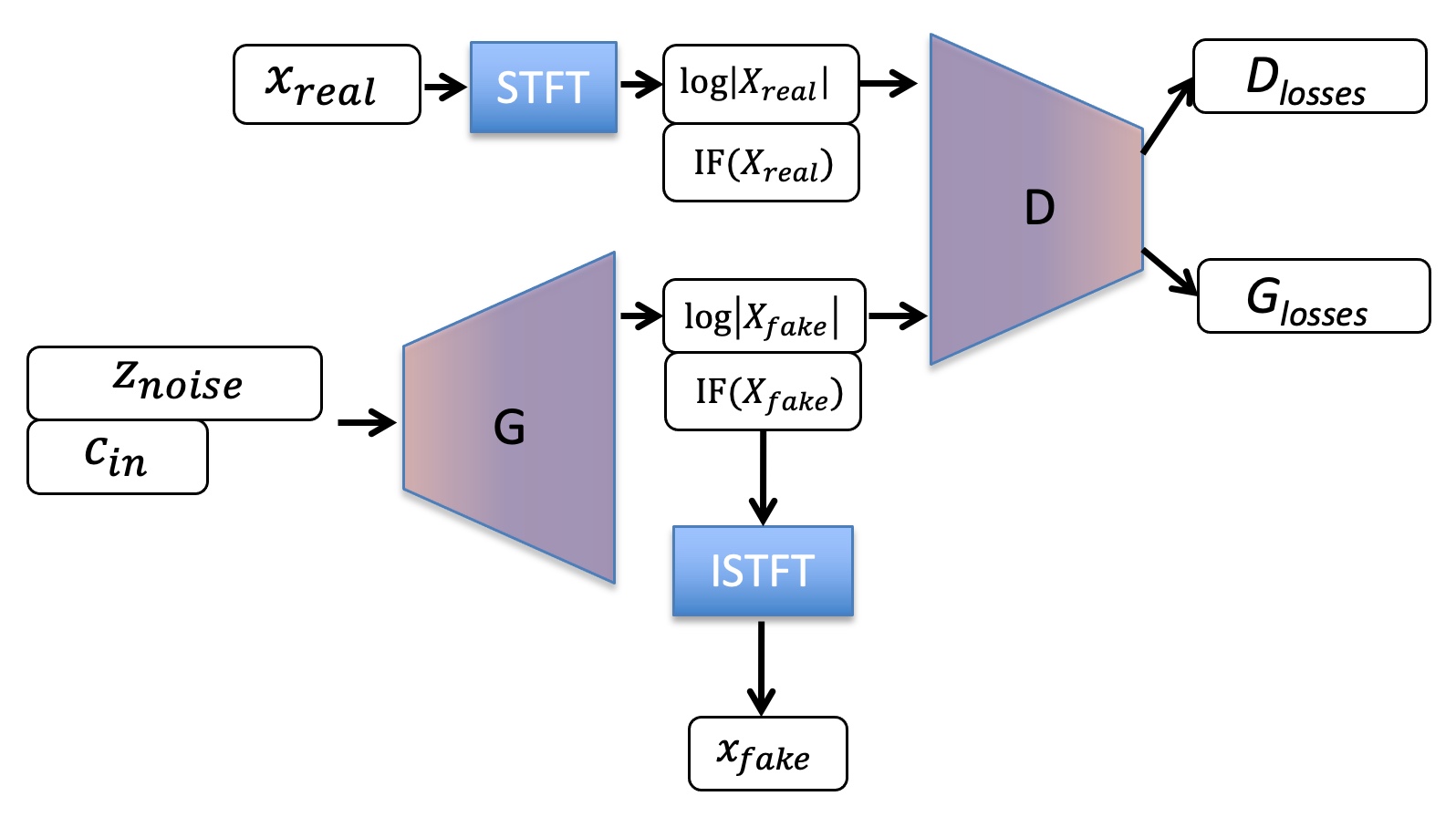}}\quad
    \subfloat[]{\includegraphics[width=\linewidth]{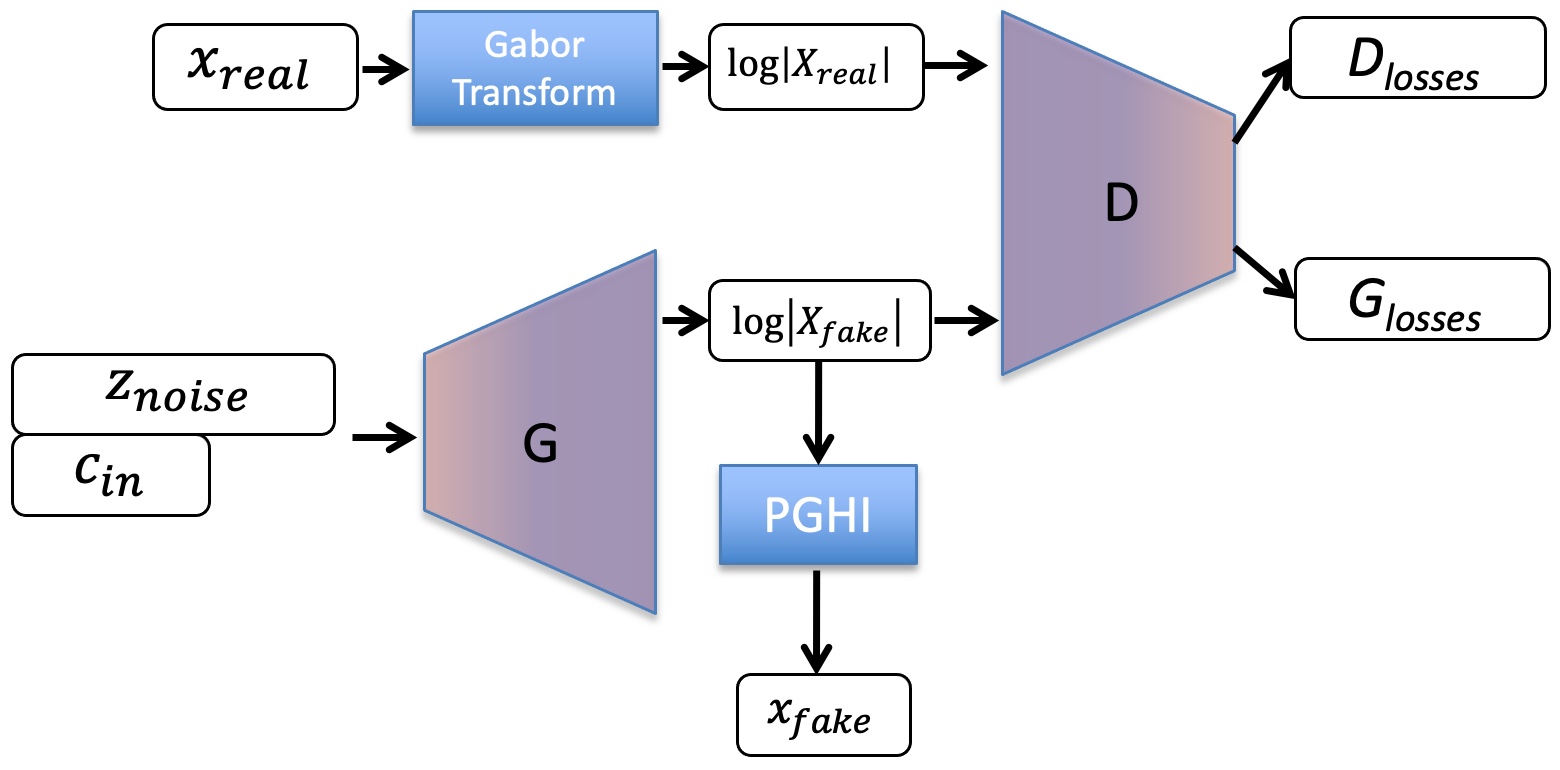}}\quad
	    \caption{GAN block diagram with (a) IF, and (b) PGHI. $z_{noise}$ is the 128 dimensional latent vector, $c_{in}$ is the conditional parameter one-hot vector. G is the generator, D is the discriminator.}
	    \label{fig:blockdiagram}
	\end{figure}

	We used the progressively growing Wasserstein GAN architecture \cite{nistal2021comparing,engel2019gansynth} which consists of a generator G and a discriminator D, where the input to G is a random vector $z$ with 128 components from a spherical Gaussian along with a one-hot conditional vector $c_{in}$. Separate models were trained for each data set with the only difference being the dimension of the one-hot pitch vector (27, 13, and 9 for NSynth, pops, and chirps, resp.)  
	For each dataset, we train two models as shown in Figure \ref{fig:blockdiagram}. Model A uses a 2-channel audio representation consisting of the log magnitude spectrogram and IF (Figure \ref{fig:blockdiagram}(a)) computed from Short Time Fourier Transform (STFT) with Hanning window, and Model B uses a single-channel log magnitude of Gabor transform (i.e.~STFT with Gaussian window) audio representation (Figure \ref{fig:blockdiagram}(b)). During generation, Model A's estimated IFSpectrogram is inverted to a real time domain signal using Librosa's inverse STFT which uses Griffin-Lim iterative algorithm for synthesis initialized by the estimated phase from IF. For model B, we use phase gradient heap integration (PGHI) \cite{pruuvsa2017noniterative}\footnote{https://github.com/andimarafioti/tifresi} for reconstruction of the audio signal from the log magnitude. It reconstructs the phase only for the positive frequency coefficients and enforces conjugate symmetry to the negative frequency coefficients in order to guarantee a real-valued time domain signal.

	The generator’s architecture consists of a Format block and a stack of Scale blocks. The Format block turns the 1D input vector $z$ + one hot conditional $c_{in}$, with 128 + x dimensions (where x could be 27, 13, or 9) into a 4D convolutional input consisting of [batch size, 128, $w_0$, $h_0$], where $w_0$ and $h_0$ are the sizes of each dimension at the input of the scale block.
	
	The scale blocks are a stack of convolutional and box-up-sampling blocks that transform the convolutional input to the generated output signal progressively in 5 phases. The discriminator D is composed of convolutional and down sampling blocks, mirroring the configuration of the generator. D estimates the Wasserstein distance between the real and generated distributions. For more details, please refer to \cite{nistal2021comparing}\footnote{https://github.com/SonyCSLParis/Comparing-Representations-for-Audio-Synthesis-using-GANs}.
	Our code that implements the GAN architecture with IF as well as PGHI methods (an extended version of Nistal et al.'s code) is available here: \url{https://github.com/lonce/sonyGanFork}. 
	
	\subsubsection{Training}
	Training is divided into 5 phases, wherein each phase a new layer, generating a higher-resolution output, is added to the existing stack, which is the essence of the progressive-GAN \cite{engel2019gansynth,karras2017progressive}. The gradual blending in of the new layers ensures minimum possible perturbation effects as well as stable training. We train all the models for 1.2M iterations on batches of 8 samples: 200k iterations in each of the first three phases and 300k in the last two. Adam optimization method is employed.
	
	Time-frequency representations of 16kHz sampled audio are computed using an FFT size of 512. We tested the effect of redundancy between frames in reconstruction, thus we trained two models, with hop sizes 64 and 128, i.e.~87.5\% and 75\% overlap between consecutive frames. We train two types of models IF and PGHI, for three kinds of audio textures, NSynth, pop, and chirp, for each of the two hop sizes.
	\subsection{Evaluation Metrics}
	Evaluation of generative models is challenging, especially when the goal is to generate perceptually realistic audio that may not be exactly same as any real audio in the dataset. Previously, the inception score has been used as the objective measure that evaluates the performance of a model for a classification task such as pitch or instrument inception score \cite{engel2019gansynth,nistal2021comparing}. However, in this work, we are comparing signal representations and synthesis techniques, while the GAN architecture remains the same. Since the variety of sounds with respect to classification is not expected to change. Indeed, Nistal et al\cite{nistal2021comparing} noted that inception models are not robust to the particular artifacts of the representations they were comparing, and therefore, it is not a very reliable measure of the overall generation quality.
	
	Marafioti et al.~\cite{marafioti2019adversarial}
 developed an interesting \textit{consistency} measure that estimates how close a magnitude spectrogram is to the frequency transform of a real audio signal. However, it is not obvious how it could be used to compare representations that include explicit phase representations. Also, the perceptual quality of the generated audio signal depends on other factors as well. For example, a real-valued time domain signal of poor perceptual quality will  have a perfectly consistent magnitude spectrogram.
 
 In this work, we performed listening tests for subjectively evaluating the quality of the generated sounds, as well as computed Fr\'echet Audio Distance (FAD) \cite{kilgour2018fr} as the objective evaluation metric. FAD is a measure of the statistical distance between real audio and fake audio from a trained model, which has been found to have some correlation with human perceptual judgment.
	\subsubsection{Human Evaluation}
	To construct stimuli for listening experiments, three points in the latent space are randomly chosen to generate three audio signals of 1 second each per pitch class per trained model, which were then stitched together with a 0.5 second silence before each of the 3 segments) resulting in a 4.5 seconds duration audio clips that were presented in the listening test. This provided variability within each clip so that the listeners focus on the sound quality of the clips and not on the instrument type or the rate of pops and chirps. For reference, a similar set of audio clips was prepared from the original or real audio data set as well. 
	
	The listening test was conducted by recruiting twenty participants via Amazon's Mechanical Turk (AMT) website. In each assessment task, the participants were asked to listen first to the reference, then to the two synthesized audio clips,  randomly ordered, and then to select the one they felt was the closest in sound quality to the reference clip, or if they were similar. The two audio clips belonged to either IF or PGHI reconstruction techniques for a hop size of 64 or 128 for each comparison. Only same type of sounds were compared, i.e. NSynth\_IF to NSynth\_PGHI, pop\_IF to pop\_PGHI etc. Moreover, the two clips being compared had the same pitch or center frequency. 20 random pitches from the NSynth dataset, 13 pitches from pops, and 9 pitches from chirps were selected to build a sample size of 84 comparison trials (42 comparisons each for hop 64 and 128 reconstructions respectively) and overall 1,680 ratings were collected. The trials were loaded into AMT in a random sequence and were completed by participants within 2 hours. The participants were compensated at the rate of US\$ 0.02 per comparison trial.
	\subsubsection{Fr\'echet Audio Distance}
	The Frechet Audio Distance (FAD) \cite{kilgour2018fr}\footnote{https://github.com/google-research/google-research/tree/master/frechet\_audio\_distance} is the distance between the statistics (mean and covariance) of real and fake data computed from an embedding layer of the pre-trained VGGish model. The embedding layer is considered to be a continuous multivariate Gaussian, where the mean and covariance are estimated for real and fake data, and the FAD between these is calculated as:
	\begin{equation}
	    FAD=||\mu_r-\mu_g||^2+tr(\Sigma_r+\mu_g-2\sqrt{\Sigma_r\Sigma_g})
	\end{equation}
where $\mu_r,\Sigma_r$ and $\mu_g,\Sigma_g$ are the mean and covariances of real and fake probability distributions, respectively. Lower FAD means smaller distances between synthetic and real data distributions. The VGGish model is trained on 8M Youtube music videos with 3K classes. The FAD metric has been tested successfully specifically for the purpose of reference-free evaluation metric for enhancement algorithms. FAD performs well in terms of robustness against noise, computational efficiency, and consistency with human judgments, and has been used by Nistal et al.\cite{nistal2021comparing}.

    \subsection{Results and Discussion}
	Qualitatively it is observed that with the IF method, the sharp transients of the pop sounds get smeared in time, whereas PGHI method produces clear and sharp transients. This temporal smearing effect is also observed in the short duration chirps generated from the IF method. This smearing effect arises from the inability of IF to provide robust information about phase when the signal contains closely spaced wideband frequency components. For NSynth data, however, the two methods sounded approximately equal in quality. Examples of the synthesised audio presented for listening tests are here: \url{https://animatedsound.com/amt/listening_test_samples/#examples}, and visual analysis of the generated spectrograms are provided here: \url{https://animatedsound.com/amt/listening_test_samples/#analysis}.
	
\begin{figure}[h]
	    \centering
        \includegraphics[width=\linewidth]{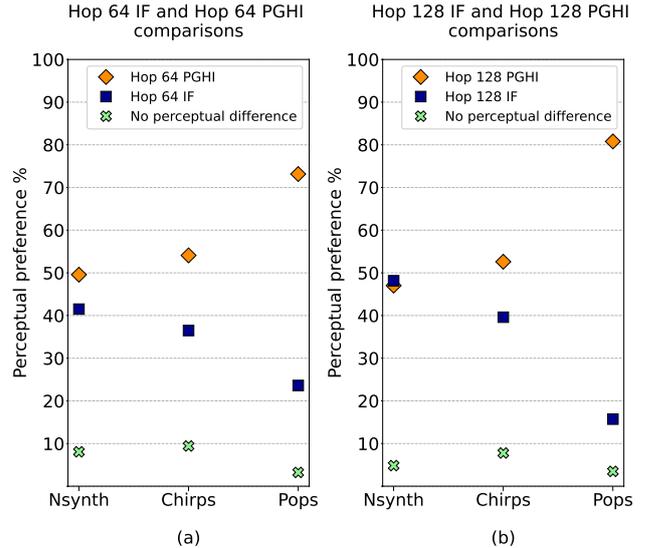}
        \caption{Results from listening tests for comparing IF and PGHI reconstructions from GAN using hop lengths of (a) 64 and (b) 128 respectively. Across both hop lengths, PGHI reconstructions of noise bursts or pops were rated to be significantly better than IF. For chirps, PGHI reconstructions were rated to be slightly better than IF and for pitched instruments PGHI reconstructions were rated almost similar to IF. }
        
    	\label{fig:listeningtestplot}
	\end{figure}

	Figure \ref{fig:listeningtestplot} (a) and (b) show results from the listening test for reconstructions using hop sizes 64 and 128 respectively. For both hop sizes, participants rated PGHI reconstructions to be significantly better than IF for pop sounds, where they rated in favour of PGHI 80.79\% and 73.15\% for hop sizes 128 and 64 respectively. This result clearly shows that PGHI with GAN produces perceptually higher quality audio for noisy signals. For chirp sounds, participants rated PGHI somewhat better than IF. But for NSynth pitched instrument sounds, PGHI and IF are similarly rated for both hop lengths. Furthermore, we observe that hop size 64 shows a clearer distinction in preference between IF and PGHI for nsynth and chirp sounds, than hop size 128. This indicates that a higher redundancy in the spectrogram representation may help in better reconstruction with PGHI method than IF method. However, comparison between the two hop sizes for the same method has shown mixed responses for the different datasets, which means that redundancy of more than 4 may not have a significant impact on the reconstructed audio quality of one method.
	This systematic study suggests that PGHI with GAN produces audio quality perceived as roughly equal to the state-of-the-art IF method for pitched sounds, but significantly higher as the complexity of the signal increases.
	
	To evaluate objectively, we computed the FAD metric, as shown in Table \ref{tab:fad}. We observe that PGHI method generated audio that consistently shows a smaller distance from reference audio compared to that generated from IF method, although unlike the perceptual ratings, the two representations are closer for chirps than the other two signal types. While this objective measure is broadly in line with the higher ratings for the PGHI method, the systematic disagreement between the user and objective measures across pitched and chirp sounds demonstrate that there is more work to be done to find an objective measure that correlates with human judgements of quality. 

	\begin{table}[]
\centering
\resizebox{0.7\columnwidth}{!}{%
\begin{tabular}{|l|l|l|l|}
\hline
 \textbf{Audio Texture} & \textbf{Hop Size} & \textbf{IF} & \textbf{PGHI} \\ \hline
Pitched Instruments & 128 & 1.500 & 1.001 \\ \hline
Pitched Instruments & 64 & 1.583 & 0.924 \\ \hline
Pops & 128 & 1.783 & 0.305 \\ \hline
Pops & 64 & 1.866 & 0.295 \\ \hline
Chirps & 128 & 1.395 & 1.031 \\ \hline
Chirps & 64 & 1.269 & 0.747 \\ \hline
\end{tabular}%
}
\caption{FAD results of different GAN models with IF and PGHI}
\label{tab:fad}
\end{table}

\section{Discussion}
We present a general method of audio synthesis using GAN that produces high quality audio output for a wide variety of sounds, pitched instruments as well as non-pitched and noisy pop and chirp sounds. We show that IFSpectrogram representation that currently produces the state-of-the-art performance with GAN for pitched instruments is not a robust representation for non-pitched and noisy sounds. Moreover, through subjective and objective measures, we show that integrating the PGHI representation and reconstruction technique in the GAN framework provides a reasonable solution to this problem, as it generates better audio quality for noisy pops and chirps than when using the IFSpectrogram method, and produces similar audio quality for pitched instruments. Audio examples generated from our experiments are available here: \url{https://animatedsound.com/amt/listening_test_samples/}, and our code implementation is available here: \url{https://github.com/lonce/sonyGanFork}.

A potential direction of improvement of the PGHI technique is to use the phase estimates from PGHI as a \textit{warm-start} for other iterative phase reconstruction algorithms such as LeGLA, as shown by Prusa et al.\cite{pruuvsa2017noniterative}. Another possibility is to include different explicit representations of phase information in training that might outperform magnitude-only reconstruction with PGHI. Marafioti \cite{marafioti2019adversarial} used a representation with frequency derivatives for training which did not perform as well as the magnitude PGHI reconstruction method, but indicates the potential that this direction has to offer.

  The method of training a GAN as a data-driven approach to designing parametrically controlled synthesizers holds a lot of promise for creative applications such sound design and music. A signal-independent representation for training the networks is an important step towards the universality and usability of this approach.







\color{black}

	\begin{acknowledgments}
This research is supported by a Singapore MOE Tier 2 grant MOE2018-T2-2-127, and by an NVIDIA Corporation Academic Programs GPU equipment grant.
	\end{acknowledgments} 

	\bibliography{smc2021bib}
	
\end{document}